\documentclass{llncs} 

\usepackage{url}
\usepackage{cite}
\usepackage{amsmath,amssymb,amsfonts}
\usepackage{algorithmic}
\usepackage{graphicx}
\usepackage{textcomp}

\usepackage{amsmath}
\usepackage{amssymb}
\usepackage{amsfonts}
\usepackage{verbatim}
\usepackage{float}
\usepackage{todonotes}
\usepackage{latexsym}
\usepackage{alltt}

\usepackage{enumerate}
\usepackage{booktabs}
\usepackage{multirow}

\usepackage{colortbl}

\usepackage{booktabs}

\newcolumntype{P}[1]{>{\centering\arraybackslash}p{#1}}

\usepackage{url}

\long\def\comment#1{}

\newcommand{\caI}{{\cal I}}

\newcommand{\caP}{{\cal P}}

\newcommand{\caX}{{\cal X}}

\newcommand{\sort}[1]{\ensuremath{\mathsf{#1}}}

\newcommand{\Variables}{\caX}
\newcommand{\Symbols}{\Sigma}

\newcommand{\nI}[1]{\ensuremath{#1{\:\notin\:}\caI}}
\newcommand{\inI}[1]{\ensuremath{#1{ \:\in\: }\caI}}

\newcommand{\NPASymbols}{\Symbols_\mathcal{P}}
\newcommand{\NPAEq}{E_\mathcal{P}}

\newcommand{\SSEq}{E_\mathit{SS}}
\newcommand{\SSSymbols}{\Symbols_\mathit{{SS}}}

\begin{document}

\title{Formal verification of the YubiKey and YubiHSM APIs in  Maude-NPA }

\author{
Antonio Gonz{\'a}lez-Burgue\~no\inst{1}
\and
Dami{\'a}n Aparicio\inst{2}
\and
Santiago Escobar\inst{2}
\and 
Catherine Meadows\inst{3}
\and 
Jos{\'e} Meseguer\inst{4}
}

\institute{
University of Oslo, Norway
\email{antonigo@ifi.uio.no}
\and
DSIC-ELP, Universitat Polit\`ecnica de Val\`encia, Spain \\
\email{\{daapsnc@dsic.upv.es, sescobar@dsic.upv.es\}}
\and
Naval Research Laboratory, Washington DC, USA 
\email{meadows@itd.nrl.navy.mil}
\and
University of Illinois at Urbana-Champaign, USA 
\email{meseguer@illinois.edu}
}

\maketitle

\begin{abstract}
In this paper, we perform an automated analysis of two devices developed by Yubico: \emph{YubiKey}, designed to authenticate a user to network-based services, and \emph{YubiHSM}, Yubico's hardware security module.
Both are analyzed using the Maude-NPA 
cryptographic protocol analyzer. 
Although previous work has been done applying automated tools to these
devices, to the best of our knowledge there has been no completely automated analysis  to date.  This is not
surprising, because both YubiKey 
and YubiHSM, which make use of cryptographic APIs, involve a number of complex features:  
(i) discrete time in the form of \emph{Lamport clocks},
(ii) a mutable memory for storing previously seen keys or nonces, 
(iii) event-based properties that require an analysis of sequences of actions, and 
(iv) reasoning modulo exclusive-or. 
In this work, we have been able to both prove properties of YubiKey 
and find the known attacks on the 
YubiHSM, in a completely automated way  beyond the capabilities of previous work in the literature.
\end{abstract}

\pagestyle{plain}

\section{Introduction}


Nowadays there exist several security tokens having the form of a smartcard or  an USB device, which are designed for protecting cryptographic values from an intruder, e.g, hosting service, email, e-commerce, online banks, etc. They are also used to ease authentication for the authorized users at a service, e.g., if you are using a service that verifies your Personal Identification Number (PIN), the same service  should not be used for checking your flights, reading your emails, etc. 
By using an Application Programming Interface (API) to separate the service from the authenticator system, 
such problems can be prevented.

Yubico is a leading company on open authentication standards and has developed two core inventions: the \emph{YubiKey}, a small USB designed to authenticate a user against network-based services, and the \emph{YubiHSM}, Yubico's hardware security module (HSM). 
The YubiKey allows for the secure authentication of a user against network-based services
by considering different methods: one-time password (OTP), public key encryption, public key authentication, and the Universal 2nd Factor (U2F) protocol \cite{FIDO}.
YubiKey works by using a secret value (i.e., a running counter) and some random values, all encrypted using  a 128 bit Advanced Encryption Standard (AES). An important feature of YubiKey is that it is independent of the operating system and does not require any installation, because it works with the USB system drivers.
YubiHSM is intended to operate in conjunction with a host application.
It 
supports several modes of operation, but the key concept is a symmetric scheme where one device at one location can generate a secure data element in a secure environment. Although the main application area is for securing YubiKey's OTP authentication/validation operations, the use of several generic cryptographic primitives allows a wider range of applications.
The increasing success of
YubiKey and YubiHSM
has led to its use by governments, universities and companies like Google, Facebook, Dropbox, CERN, Bank of America etc., including more than 30,000 customers \cite{Yubico07}.

Cryptographic Application Programmer Interfaces (Crypto APIs) are commonly used to secure interaction between applications and hardware security module (HSMs), and are both used in YubiKey and YubiHSM. However, many crypto APIs have been subjected to intruder manipulation to disclose relevant information, as is the case for YubiHSM.
In \cite{Kunnemann12,Kunnemann14}, K{\"u}nnemann and Steel show two kinds of attacks on the first released version YubiHSM API:
(i) if the intruder had access to the server running YubiKey, where AES keys are generated, then it was able to obtain plaintext in the clear;
(ii) even if the intruder had no access to the server running YubiKey, it could use previous nonces  to obtain AES keys.

This paper is the third in a series using Maude-NPA to analyze cryptographic APIs; earlier work
appeared in \cite{ssr2014,ssr2015}.  We find this problem area one of particular interest for two reasons. First, these  APIs often use exclusive-or and this gives us the opportunity to explore how well Maude-NPA, which
provides support for the full exclusive-or theory, can be applied to protocols that use exclusive-or.
Secondly, cryptographic APIs offer a number of other challenging features such as mutable global
state, and this allows us to explore how Maude-NPA can be made to handle global state as well.

In this work we 
use Maude-NPA~\cite{Maude-NPA} 
for analyzing both YubiKey and YubiHSM.
Our analysis was carried out on generation 2 of YubiKey and version 0.9.8 beta of the YubiHSM, as was the analysis of \cite{Kunnemann14}.  In order to facilitate comparison with earlier work, our formal specifications of YubiKey
and YubiHSM follow those of \cite{Kunnemann14} as closely as possible.
Our main contributions are:
\begin{enumerate}
\item We have been able 
to prove secrecy and authentication properties of YubiKey and to find both attacks on YubiHSM, 
beyond the capabilities of the earlier analysis, which was only able to find one, due to limited
support for exclusive-or.
To the best of our knowledge,
ours is the only analysis of YubiKey and YubiHSM
using a protocol analysis tool with a full treatment of exclusive-or.
\item  
The analysis was completely automatic and either found an attack or terminated with
a finite search graph, showing that no attack of that kind exists. 
That is, Maude-NPA did not need any human guiding or auxiliary lemmas.
The earlier analysis involved 
some additional user-defined lemmas 
in order to prove one of the
properties of YubiKey.
\item We show in Section~\ref{sec:maude-npa}  how we implemented
(i) Lamport clocks, (ii) mutable memory, and (iii) event-based properties in Maude-NPA,
even though the tool does not support these natively,
by making use of Maude-NPA's capabilities for protocol composition and reasoning modulo associativity.
These techniques should be applicable to other protocols with similar properties.
\end{enumerate}

\section{The YubiKey Device}
\label{intro-yubikey}

The  YubiKey USB device \cite{Yubico-manual} is an authentication device capable of generating One Time Passwords (OTPs). The YubiKey connects to a USB port and identifies itself as a standard USB keyboard, which allows it to be used in most computing environments using the system's native drivers.

 We will focus on the YubiKey OTP mode, a mode that uses a button physically located on the YubiKey. When this button is pressed, it emits a string that can be verified only once against a server in order to receive the permission to access a service. Furthermore, a request for a new authentication token is triggered also by touching the YubiKey button. As a result of this request, some counters that are stored on the device are incremented and some random values are generated in order to create a fresh 16-byte plaintext. 
 An OTP has the following concatenated fields \cite{Vamanu12}:

\begin{center}
\includegraphics[width=.6\textwidth]{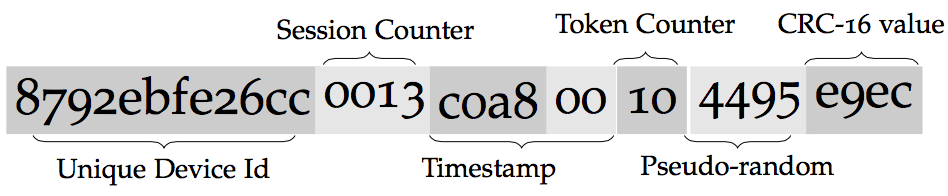}
\end{center}

The YubiKey authentication server accepts an OTP only if it decrypts under the correct AES key into a valid secret value containing a counter larger than
the last accepted counter. 
The token counter is used as a \emph{Lamport clock} \cite{Lamport:1978:TCO:359545.359563}
to determine the order of events in a distributed concurrent system.

\begin{figure}[htb]
\begin{center}
\includegraphics[width=0.6\textwidth]{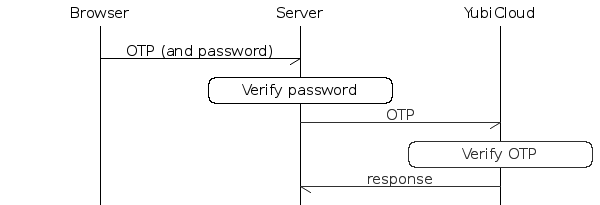}\centering \end{center}
\caption{YubiKey OTP Validation Flow}
\label{fig:validation-flow} 
\end{figure}

The authentication protocol of YubiKey involves three roles: (i) the user, (ii) the service, and (iii) the verification server. The user can have access to the service if it provides its own valid OTP generated by the YubiKey; its validity is verified by the verification server as explained before. Figure \ref{fig:validation-flow} \cite{Merkel09} is a simple example of a YubiKey API execution, 
where the three roles are as follows:
the user (Browser), the service (YubiCloud), and the verification server running the YubiKey API.

The YubiKey OTP generation scheme can be described by the following interaction sequence.
\begin{enumerate}
\item The initialization of the YubiKey device takes place. 
A fresh public ID (\textit{pid}), secret ID (\textit{sid}) and YubiKey key (\textit{k}) are generated. 
	Any interaction between the YubiKey and the server will involve all three elements \textit{pid}, \textit{sid} and \textit{k}.
	There are also two token counters, one stored on the Server and another stored on the YubiKey. 
	The server saves information using the notation \texttt{SharedKey(pid,k)} and \texttt{Server(pid,sid,token counter)},
	whereas
	the YubiKey saves information using \texttt{Y(pid,sid)} and \texttt{YubiCounter(pid,token counter)}.
\item The YubiKey is plugged in. 
	Every time the YubiKey is plugged in, the YubiKey token counter must be increased.
	However, we consider a compromised scenario in which the intruder could produce all counter values as in \cite{Kunnemann14}, thus
	adding a new token counter as an input to the command and checking that it must be
	bigger than the old stored token counter.
	Figure~\ref{fig:yubi-plugin} shows a graphical representation
	of the plugin event, including the input, output, and saved information.

\begin{figure}[h]
\begin{center}
\includegraphics[width=.7\textwidth]{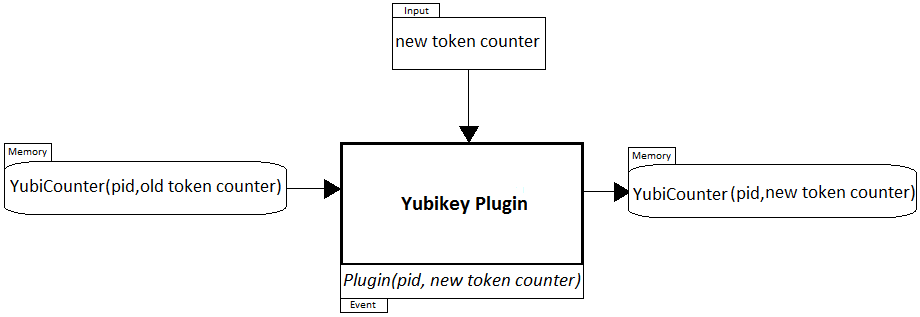}\centering \end{center}
\caption{YubiKey Plugin API Command}
\label{fig:yubi-plugin}  
\end{figure}

	\item The user pushes the YubiKey OTP generation button and generates a byte string formed by the \textit{sid}, the YubiKey token counter, 
	and a random number. 
	The byte string is encrypted using a symmetric encryption operator and the saved key \textit{k}. 
	The YubiKey token counter is also increased.
	According to the compromised scenario, the YubiKey token counter must be provided as input.
	Figure~\ref{fig:yubi-press} shows a graphical representation
	of the button-pressing event, including the input, output, and saved information.
	
\begin{figure}[h]
\begin{center}
\includegraphics[width=.8\textwidth]{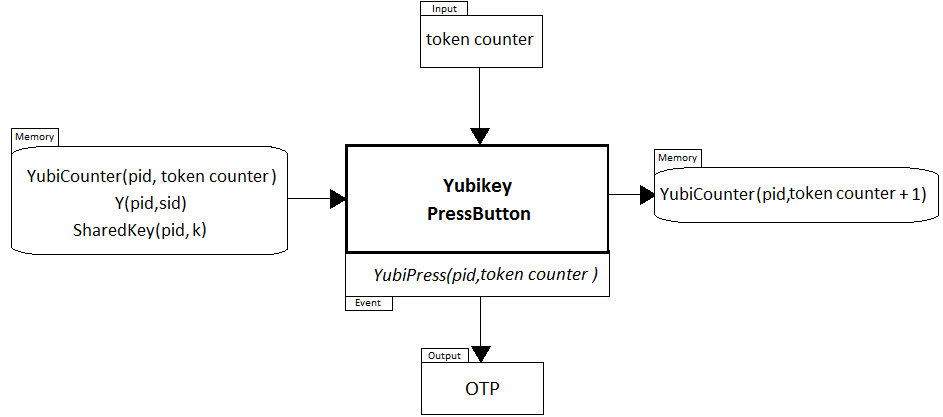}\centering \end{center}  
\caption{YubiKey Press Button Command}
\label{fig:yubi-press} 
\end{figure}

	\item 
	Upon reception of the generated OTP string, the basic verification steps are: 
\begin{enumerate}[4.1] 
		\item The byte string is decrypted, and if it is not valid the OTP is rejected. 
		\item The token counter stored in the OTP is compared with the server token counter. 
		If smaller than or equal to the server token counter, the received OTP is rejected as a replay.
		According to the compromised scenario, the server token counter must be provided as input.
		\item 
		A successful login must have been preceded by a button press for the same counter value, and there is not a second distinct login for this counter value. In this
		paper we omit this check and show that this property is always guaranteed, assuming that the checks on the byte string and token counter succeed.  
		\item If all the checks succeed, 
		the token counter stored in the OTP is stored as the server token counter and the OTP is accepted as valid.
	\end{enumerate}
	Figure~\ref{fig:yubi-login} shows a graphical representation
	of the login event, including the input, output, and saved information.

\begin{figure}[hbt]
\begin{center}
\includegraphics[width=.9\textwidth]{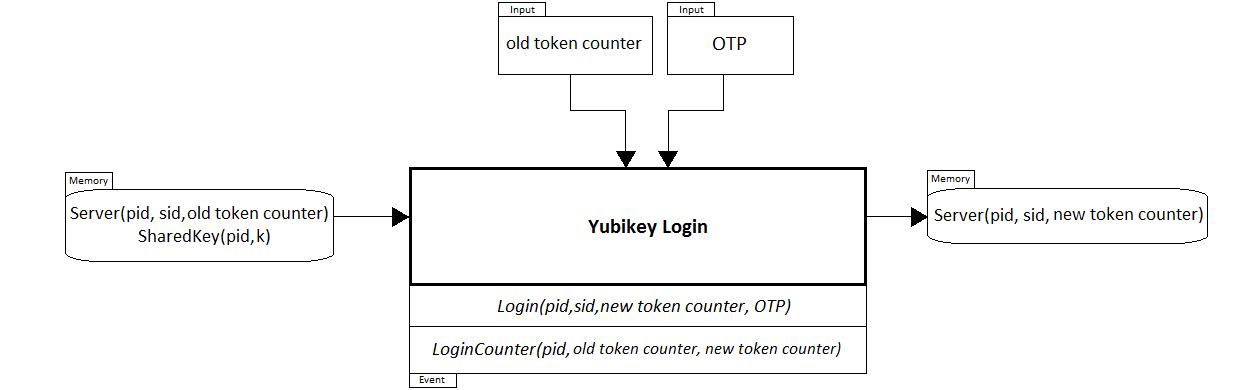}\centering \end{center}  
\caption{YubiKey Login Command}
\label{fig:yubi-login} 
\end{figure}
	
\end{enumerate}

\noindent
In \cite{Kunnemann12,Kunnemann14}, K{\"u}nnemann and Steel 
were able to prove several properties:
\begin{enumerate}[(a)]
\item Absence of replay attacks, i.e., there are no two distinct logins that accept the same counter value. 
\item Correspondence between pressing the button on a YubiKey and a successful login. 
In other words, a successful login must have been preceded by a button pressed for the same counter value. Furthermore, there is no second distinct login for this counter value. 
\item Counter values are different over time, where a successful login invalidates previous \textit{OTP}s. 
Note that the verification of this property in Tamarin involved 
additional user-defined lemmas.

\end{enumerate}

\section{The YubiHSM Device}
\label{intro-yubihsm}

Yubico also distributes a USB device that works as 
an application-specific Hardware Security Module (HSM)
to protect the YubiKey AES keys. 
The YubiHSM~\cite{YubiHMS-manual} stores a very limited number of AES keys in a way that the server can use them to perform cryptographic operations without the key values ever appearing in the server's memory.  The YubiHSM is designed to protect the YubiKey AES keys when an authentication server is compromised by encrypting the AES keys using a master key stored inside the YubiHSM.

In addition, the YubiHSM can decrypt an indefinite number of YubiKey's OTP's  with secure storage of the AES keys on the host computer. 
The AES keys are only readable to the YubiHSM through the use of \textit{Authenticated Encryption with Associated Data (AEAD)}. The AEAD uses a cryptographic method that provides both confidentiality and authenticity. 
An AEAD consists of two parts: (i) the encryption of a message using the counter mode cryptographic mode of operation,   and (ii) a \textit{message authentication code (MAC)} taken over the encrypted message. In order to construct, decrypt or verify an AEAD, a symmetrical cryptographic key and a piece of associated data are required. This associated data, called a \emph{nonce} in the rest of the paper,
 can either be a uniquely generated handle or something that is uniquely related to the AEAD.

To encrypt a message using counter mode, one first divides it into blocks of equal length, each suitable for input to the block cipher AES, e.g. $\mathit{data_1},  \ldots , \mathit{data_n}$.
The sequence $\mathit{counter_1}, \dots, \mathit{counter_n}$ is then computed, where $\mathit{counter_i} = \mathit{nonce} \oplus i$ modulo $2^\eta$ where $\eta$ is the
length of a block in bits. The encrypted message is then $\mathit{senc}(\mathit{counter_1},k) \oplus \mathit{data_1} ; \ldots ; \mathit{senc}(\mathit{counter_n},k) \oplus \mathit{data_n}$,
where $\mathit{senc}$ is the encryption function and $k$ the symmetrical cryptographic key, and 
     \linebreak 
$\mathit{senc}(\mathit{counter_1},k)  ; \ldots ; \mathit{senc}(\mathit{counter_n},k)$  is called the
\textit{keystream}. Finally, the MAC is computed over the encrypted message and appended to obtain
     \linebreak 
$(\mathit{senc}(\mathit{counter_1},k) \oplus \mathit{data_1} ; \ldots ; \mathit{senc}(\mathit{counter_n},k) \oplus \mathit{data_n}) ; \mathit{MAC}$.  The MAC is of fixed length,
so it is possible to predict where it starts in an AEAD.
However, since the two attacks considered below do not involve most of the details about block cipher AES,
we follow the generalization of \cite{Kunnemann14}
and consider just messages of the form
$\mathit{senc}(\mathit{cmode}(\mathit{nonce}),k) \oplus \mathit{data} ; \mathit{mac}(\mathit{data},k)$.

In \cite{Kunnemann12,Kunnemann14}, K{\"u}nnemann and Steel reported two kinds of attacks on version 0.9.8 beta of YubiHSM API:
(a) if the intruder has access to the sever running YubiKey, where AES keys are generated, then it is able to obtain plaintext in the clear;
(b) even if the intruder has no access to the server running YubiKey, it can use previous nonces  to obtain AES keys.

\begin{figure}[h]
\begin{center}
\vspace{-1ex}
\includegraphics[width=.4\textwidth]{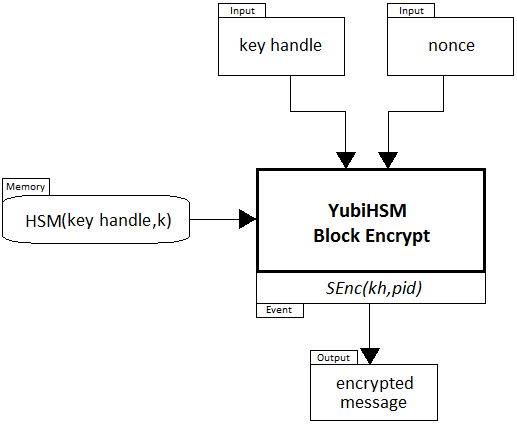}\centering \end{center}
\vspace{-1ex}
\caption{YubiHSM Block Encrypt API Command}
\label{fig:yubiHSM-block-encrypt} 
\vspace{-1ex}
\end{figure}

The first attack
involves the YubiHSM API command 
depicted in Figure \ref{fig:yubiHSM-block-encrypt}, which takes a handle to an AES key and 
the nonce 
and applies the raw block cipher. 

In order to perform this attack, the intruder compromises the server to learn a \textit{AEAD} and the key-handle used to produce it.
Then, using the Block Encrypt command
shown in Figure
 \ref{fig:yubiHSM-block-encrypt}, an intruder is able to decrypt an \textit{AEAD} by recreating the blocks of the key-stream: inputting $\mathit{counter_i}$ (the nonce) to the YubiHSM
Block Encrypt API command.   The intruder exclusive-ors the result with the \textit{AEAD} truncated by the length of the \textit{MAC}  and obtains the plaintext. 

\begin{figure}[h]
\begin{center}  
\includegraphics[width=.5\textwidth]{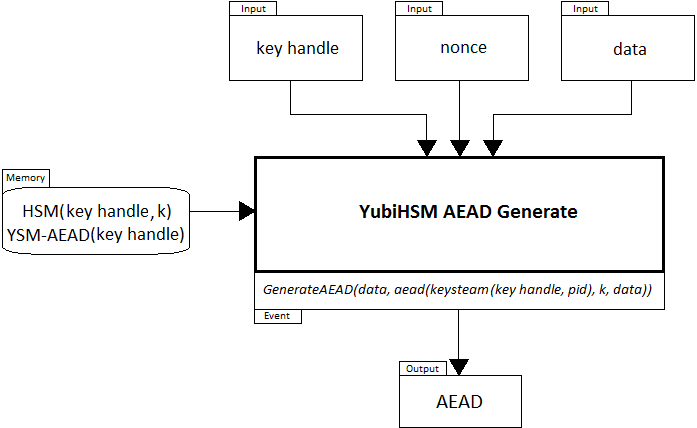}\centering \end{center}
\vspace{-3ex}
\caption{YubiHSM AEAD Generate API Command}
\label{fig:yubiHSM-aead-generate} 
\vspace{-2ex}
\end{figure}

The second attack 
involves the 
YubiHSM command depicted in 
Figure~\ref{fig:yubiHSM-aead-generate}
that takes a nonce, a handle to an AES key and some data and outputs an \textit{AEAD}. An intruder can produce an \textit{AEAD} for the same handle \textit{kh} and a value \textit{nonce} that was previously used to generated another \textit{AEAD}. An intruder can recover the keystream directly  by using the AEAD-Generate command to encrypt a string of zeros and discarding  the \textit{MAC}. The result will be the exclusive-or of the keystream with a string of zeros, which is equal to the keystream itself. This attack is worse than the first one, because this command cannot be avoided or restricted (see \cite{Kunnemann14} for further details). 

\section{Maude-NPA}\label{sec:maude-npa}
 
 In Maude-NPA, as in most formal analysis tools for cryptographic protocols,
 a protocol is a set of rules that describe the actions of honest principals communication
 across a network controlled by an intruder.  Given a protocol $\mathcal{P}$,
states in Maude-NPA are modeled as
elements of an initial algebra $\mathcal{T}_{\NPASymbols/
\NPAEq}$, where $\NPASymbols = \SSSymbols \cup \Sigma_\mathcal{C}$
 is the signature
defining the sorts and function symbols ($\Sigma_\mathcal{C}$ for the cryptographic
functions and $\SSSymbols$ for 
all the state constructor symbols),
 $\NPAEq=E_\mathcal{C} \cup \SSEq $ is a set of equations where $E_\mathcal{C}$ specifies the \emph{algebraic properties} of the cryptographic functions and $\SSEq$ denotes properties of constructors of states. The set of equations $\Sigma_\varepsilon$ may vary depending on different protocols, but the set of equations $\SSEq$ is always the same for all protocols.
Therefore, a state is an $\NPAEq$-equivalence class $[t]_{\NPAEq}\in
T_{\NPASymbols/ \NPAEq}$ with $t$ a ground
$\NPASymbols$-term, i.e. a term without variables.

In Maude-NPA 
a \emph{state pattern}
for a protocol $P$ 
is a term $t$ of sort \sort{State}
%
which has the form
$\{ S_1 \,\&\, \cdots \,\&\, S_n \,\&\, \{\textit{IK}\}\}$,
where $\&$ is an infix associative-commutative union 
operator with identity symbol $\emptyset$. 
Each element in
the set is either a \emph{strand} 
$S_i$ or the \emph{intruder knowledge} $\{\textit{IK}\}$ at that state.

The \emph{intruder knowledge} $\{\textit{IK}\}$
belongs to the state and
is represented as a set of facts 
using comma as an infix associative-commutative union operator 
with identity element $empty$. 
There are two kinds
of intruder facts: \emph{positive} knowledge facts (the intruder knows $m$,
i.e., $\inI{m}$), and \emph{negative} knowledge facts (the intruder \emph{does not
yet know} $m$ but \emph{will know it in a future state}, i.e., $\nI{m}$),
where $m$ is a message expression.

A \emph{strand}
\cite{strands} specifies the sequence of messages sent and received
by a principal executing the protocol and is represented as a sequence
of messages 
$
[\textit{msg}_1^\pm, \textit{msg}_2^\pm, \textit{msg}_3^\pm,
 \ldots,\allowbreak \textit{msg}_{k-1}^\pm, \textit{msg}_k^\pm]
 $ 
 with 
$\textit{msg}_{i}^\pm$ 
either
$\textit{msg}_i^-$ (also written\linebreak $-\textit{msg}_i$) representing an input
message, 
or
$\textit{msg}_i^+$ (also written $+\textit{msg}_i$) representing an output message.
Note that 
each $\textit{msg}_{i}$ is a term 
of a special sort \textsf{Msg}.
Variables of a special sort \sort{Fresh} are used to represent pseudo-random values (nonces)
and Maude-NPA ensures that two distinct fresh variables will never be merged.
Strands are extended with all the fresh variables $f_1,\ldots,f_k$ created by that strand,
i.e.,
$
:: f_1,\ldots,f_k:: [\textit{msg}_1^\pm, \textit{msg}_2^\pm, 
 \ldots,\allowbreak 
 \textit{msg}_k^\pm]
 $ .

Strands are used to represent both the actions of honest principals (with a strand 
specified for each protocol role) 
and the actions of an intruder 
(with a strand 
for each action an intruder is able to perform on messages).
In Maude-NPA strands evolve over time;
the symbol $|$ is used to divide past and future.
That is, given a strand 
$[\ \textit{msg}_1^{\pm},\ \ldots,\ \textit{msg}_i^{\pm}\ |\allowbreak\ \textit{msg}_{i+1}^{\pm},\ \ldots,\ \textit{msg}_k^{\pm}\ ]$,
 messages $\textit{msg}_1^\pm,\linebreak[2] \ldots, \textit{msg}_{i}^\pm$ are
the \emph{past messages}, and messages $\textit{msg}_{i+1}^\pm, \ldots,\textit{msg}_k^\pm$
are the \emph{future messages} ($\textit{msg}_{i+1}^\pm$ is the immediate future
message).
A strand 
$[\textit{msg}_1^\pm, \linebreak[2]\ldots, \linebreak[2] \textit{msg}_k^\pm]$ 
is shorthand for 
$[nil ~|~ \textit{msg}_1^\pm,\linebreak[2] \ldots,\linebreak[2] \textit{msg}_k^\pm , nil ]$.
An \emph{initial state} is a state where the bar is at the beginning for all strands in the state,
and the intruder knowledge has no fact of the form $\inI{\textit{m}}$. 
A \emph{final state} is a state where the bar is at the end for all strands in the state
and there is no intruder fact of the form $\nI{\textit{m}}$.


Since the number of 
states in
$T_{\NPASymbols/ \NPAEq}$
is in general infinite, rather than exploring concrete protocol states
$[t]_{\NPAEq}\in T_{\NPASymbols/\NPAEq}$, Maude-NPA 
explores
\emph{symbolic state patterns}  
 $[t(x_{1},\ldots,x_{n})]_{\NPAEq} \in
T_{\NPASymbols/ \NPAEq}(\Variables)$ on the free
$(\NPASymbols, \allowbreak \NPAEq)$-algebra over a set of
variables $\Variables$. In this way, a state pattern $[t(x_{1},\ldots,x_{n})]_{\NPAEq}$
represents not a single concrete state 
(i.e., an $\NPAEq$-equivalence class) but a possibly infinite set of
states
(i.e., an infinite set of $\NPAEq$-equivalence classes), namely all the \emph{instances} of the pattern
$[t(x_{1},\ldots,x_{n})]_{\NPAEq}$ where the variables $x_{1},\ldots,x_{n}$
have been instantiated by concrete ground terms. 

The semantics of Maude-NPA is expressed in terms of \emph{rewrite rules} that describe how a protocol transitions
 from one state to another 
via the intruder's interaction with it.   One uses Maude-NPA to find an attack by specifying an insecure state pattern called an \emph{attack pattern}.  Maude-NPA attempts to find a path from an initial state to the attack pattern via backwards narrowing (narrowing using the rewrite rules with the orientation reversed). 
That is, a 
narrowing sequence from an initial state to 
an attack state is 
searched \emph{in reverse} as  a \emph{backwards path} from the attack state to the initial state. 
Maude-NPA attempts to find paths until it can no longer form any backwards narrowing steps, at which point it terminates.
If  at that point it has not found an initial state, the attack pattern is judged \emph{unreachable}. 
 Note that Maude-NPA places
\emph{no bounds on the number of sessions}, so reachability is undecidable in general.  
Note also that Maude-NPA does not perform any data abstraction
such as a bounded number of nonces. 
However, the tool makes use of a number of sound and complete state space reduction techniques that help to identify unreachable and redundant states, 
and thus make termination more likely. 
 
\subsection{Modeling Mutable Memory by means of Maude-NPA Strand Compositions} 
\label{strand-comp-MNPA}

Strands can be extended with \emph{synchronization messages} \cite{DBLP:journals/corr/SantiagoEMM16} of the form 
$\{Role_1 \rightarrow Role_2 \;;;\; mode \;;;\; w\}$
where $Role_1,Role_2$ are constants of sort \textsf{Role} provided by the user,
$mode$ can be either \texttt{1-1} or \texttt{1-*} representing a one-to-one or one-to-many
synchronization (whether an output message can synchronize with one or many input messages), 
and $w$ is a term representing the information  passed
along in the synchronization messages.
Synchronization messages are limited to the beginning and/or end of a strand. 
Although originally intended for a different use,
they are very useful
for representing a strand of unspecified length as a concatenation of different fixed-length strands.
%
For example, consider a module that receives $i$ pieces of data, and
then exclusive-ors them,
i.e.,  $[ -(M_1), \ldots, -(M_i), +(M_1 \oplus \cdots \oplus M_i)]$ for $i \geq 1$.
This can be specified using three strands using synchronization messages:

\begin{align}
& 1. \ [\hspace{30.7ex} -(M_1), \{ \mathit{role}_\oplus \rightarrow \mathit{role}_\oplus \;;;\; \texttt{1-1} \;;;\; M_1 \}\hspace{7.2ex}]  \notag \\
& 2. \  [ \{ \mathit{role}_\oplus \rightarrow \mathit{role}_\oplus \;;;\; \texttt{1-1} \;;;\; M \} , -(M_2), \{ \mathit{role}_\oplus \rightarrow \mathit{role}_\oplus \;;;\; \texttt{1-1} \;;;\; (M \oplus M_2) \}] \notag \\
& 3. \  [ \{ \mathit{role}_\oplus \rightarrow \mathit{role}_\oplus \;;;\; \texttt{1-1} \;;;\; M\} , +(M)\hspace{39.5ex}] \notag
\end{align}

\noindent
Composition is then performed by unifying output synchronization messages with input synchronization messages of instances of strands.  

For the YubiKey and YubiHSM APIs, if each event is represented by a strand,
then an execution (e.g., Plugin followed by Press followed by Login)
can be represented by the concatenation of the strands associated to the execution.
However, the YubiKey and YubiHSM APIs also require different information to be stored from one API command to the next. 
Some information is read-only, but other information is updated, such as the \texttt{YubiCounter(pid,counter)}.
Maude-NPA, unlike Tamarin, does not natively support mutable memory; but it can be modeled using synchronization messages.
That is, the old data will appear in the input synchronization message of  an API strand,
and the new information will appear in the output synchronization  message of that strand,
which will then become the input synchronization message of the next API strand.



We model the mutable memory used by YubiKey as a
multiset of predicates,
where we define a new 
multiset union symbol \verb!@!,
which is an 
infix associative-commutative symbol with an identity symbol \texttt{empty}.
Thus, for the  strand describing the YubiKey button press, the input synchronization message is as follows:

\vspace{1ex}\noindent\hspace*{6ex}
{\small\texttt{\{yubikey -> yubikey ;; 1-1 ;; Y(pid,sid) @ YubiCounter(pid,c1) @ \newline
\hspace*{7ex} Server(pid,sid,c2)  @ SharedKey(pid,k)\}}}
\vspace{1ex}

\noindent
Updating the counter of the YubiKey after a button press is represented
by updating the second argument of the \texttt{YubiCounter(pid,c1)} predicate in the multiset.
This updated multiset becomes the output synchronization of the strand.


\vspace{-1ex}
\subsection{Modeling Event Lists by means of Mutable Memory} 
\vspace{-1ex}

The YubiKey and YubiHSM APIs also keep a rigid control of the ordering of events, where an event
is a state transition in the system, and a proper analysis of actions is mandatory.
Maude-NPA, unlike Tamarin, does not natively support the representation and analysis of event sequences; 
but we have implemented it by storing event sequences in the synchronization messages.
This is helped by the fact that Maude-NPA, via the Maude language, has  recently been endowed with built-in lists (using any associative  symbol).
We have defined a new infix associative symbol \verb!++! with an identity symbol \texttt{nil} to represent an event list
and also a new auxiliary infix symbol \verb!|>! where the left-hand side contains the mutable memory
and the right-hand side contains the event list. 
The input synchronization message for the button press strand 
has the form:

\vspace{1ex}
\noindent 
{\small\texttt{\{yubikey -> yubikey ;; 1-1 ;; Y(pid,sid) @ YubiCounter(pid,c1) @
                                 \newline 
                                 Server(pid,sid,c2) @ SharedKey(pid,k)
                                 |> Plugin(pid,c3) ++ Press(pid,c4)\}
}}
\vspace{1ex}

\noindent
Every time a new event occurs, it is inserted as a new element at the end of the event list.
The leftmost elements are the oldest ones, whereas the  rightmost elements are the newest. 
Thus, if we want to say that event $e_1$ must occur before event $e_2$, we can express this with
the event list $L_1\ \texttt{++}\ e_1\ \texttt{++}\ L_2\ \texttt{++}\ e_2\ \texttt{++}\ L_3$.  
,e.g., \verb+--[action facts]->+, as follows:
	[ ] 
	-->
	[ F('1','x'), F('2','y') ]

	[ F(u,v) ] 
	-->
	[ H(u), G('3',h(v)) ]
%
\verb+[ F('1','x')]+ or \verb+[F('2','y') ]+ and in this case \verb+SecondRule+ can be applied to both of them. Each of these instantiations leads to a new successor state.
%
%
%
%

\vspace{-1ex}
\subsection{Modeling Lamport Clocks in Maude-NPA Using Constraints}
\label{constraints-MNPA}
\vspace{-.5ex}

Lamport clocks require the testing of constraints: that is, whether one counter is smaller than another.
This is simple to do when the counters have concrete values. However, since Maude-NPA does not consider concrete protocol states 
but symbolic state patterns (terms with logical variables), 
the equality and disequality constraints handled by 
Maude-NPA are predicates defined over variables, whose domain, in the case
of Lamport clocks, is the natural numbers.

In Maude-NPA strands can be extended with equality and disequality constraints \cite{DBLP:conf/birthday/EscobarMMS15}
of the form $\textit{Term1}\ \texttt{eq}\ \textit{Term2}$ and $\textit{Term1}\ \texttt{neq}\ \textit{Term2}$.
Whenever an equality constraint is found during the execution of a strand,
the two terms in the equality constraint are unified 
modulo the set $E_\caP$ of equations of the protocol
and a new state is created for each possible unifier.
Whenever a disequality constraint is found during the execution of a strand,
it is simply stored in an internal repository of disequality constraints associated to each protocol state;
but every time a new state is going to be generated during the state space exploration, 
all the disequality constraints in the internal repository are tested for satisfiability \cite{DBLP:conf/birthday/EscobarMMS15}. 
That is, for each state, if there is a disequality constraint of the form $\textit{Term1}\ \texttt{neq}\ \textit{Term2}$
such that \emph{Term1} and \emph{Term2} are equal modulo $E_\caP$ then the state is discarded.



We deal with Lamport clocks symbolically by representing the relations between clocks as constraints in \emph{Presburger Arithmetic}. Although various \emph{Satisfiability modulo theories (SMT)} \cite{Nieuwenhuis:2006:SSS:1217856.1217859} solvers such as CVC4\footnote{Available at \url{http://cvc4.cs.stanford.edu/web/}.},
Yices\footnote{Available at \url{http://yices.csl.sri.com}.},
and
Microsoft Z3\footnote{Available at \url{https://github.com/Z3Prover/z3}.} could be used for this purpose, we decided to avoid the complexities of invoking an external tool while executing Maude-NPA. Instead,  we have used the variant-based decision procedure for Presburger Arithmetic already available in Maude \cite{MeseguerSCP18};
but considered only positive numbers without zero.

Adding two natural numbers $i$ and $j$ is written as $i\ \verb!+!\ j$.
Checking whether a natural number $i$ is smaller than another natural number $j$ is 
represented in Maude-NPA by a constraint of the form $j\ \verb+eq+\ i\ \verb!+!\ k$, where $k$ is a new variable.


\vspace{-1ex}
\section{Formal Specifications in Maude-NPA}\label{sec:formalModelMNPA}
\vspace{-1ex}

%

\subsection{Formal Specifications of YubiKey in Maude-NPA}\label{sec:formalModelYubiKeyMNPA}
\vspace{-.5ex}

In our specification, each command of the YubiKey API 
(Figures~\ref{fig:yubi-plugin}, \ref{fig:yubi-press}, and \ref{fig:yubi-login}) 
plus the initialization
is specified in Maude-NPA as a strand. 
	 
The initialization strand is defined as follows.
Three new \sort{Fresh} values are defined: 
a YubiKey public ID (\verb+rpid+),
a secret ID (\verb+rsid+),
and
a key `\verb+rk+' shared with the server.
Variables of sort \sort{Fresh} are wrapped by symbol \verb+Fr+ as in \cite{Kunnemann14}.

{\scriptsize\begin{alltt}
:: rk,rpid,rsid ::
[ +(init),
  \{yubikey -> yubikey ;; 1-1 ;;
   YubiCounter(Fr(rpid), 1) @ Server(Fr(rpid),Fr(rsid),1) @ 
   Y(Fr(rpid),Fr(rsid)) @ SharedKey(Fr(rpid),Fr(rk))
   |> Init(Fr(rpid),Fr(rk)) ++ ExtendedInit(Fr(rpid),Fr(rsid),Fr(rk))\}] 
\end{alltt}}

 
The API command represented in Figure \ref{fig:yubi-plugin} shows what happens when a YubiKey is being plugged in. This command checks that the new received counter is smaller than the previous one and updates the predicate \verb+YubiCounter+.

{\scriptsize\begin{alltt}
 :: nil :: 
[\{yubikey -> yubikey ;; 1-1 ;; YubiCounter(pid,otc) @ mem |> EL \},
   -(tc), (tc eq (otc + extra)),
 \{yubikey -> yubikey ;; 1-1 ;; YubiCounter(pid,tc) @ mem |> EL ++ Plugin(pid,tc)\} ] 
\end{alltt}}

\noindent
Note that the parameter \verb+mem+ denotes the rest of the mutable memory
and
the parameter \verb+EL+ denotes the previous event list.
The variable \texttt{extra} is an auxiliary variable used just for testing the numerical constraint.

The command shown in Figure \ref{fig:yubi-press} represents what happens when the YubiKey button is pressed and the OTP is sent. 
The OTP is represented by message:\newline
{\small\texttt{senc(sid ; tc ; Fr(rnpr),k)}}
where
\verb+senc+ denotes symmetric encryption using key \verb+k+
and
the infix symbol \verb+;+ denotes message concatenation.

{\scriptsize\begin{alltt}
:: rnpr,rnonce :: 
[\{yubikey -> yubikey ;; 1-1 ;;
   YubiCounter(pid,tc) @ Y(pid,sid) @ SharedKey(pid,k) @ mem |> EL \},
   -(tc),
   +(pid ; Fr(rnonce) ; senc(sid ; tc ; Fr(rnpr),k)), 
 \{yubikey -> yubikey ;; 1-1 ;;
   YubiCounter(pid,tc + 1) @ Y(pid,sid) @ SharedKey(pid,k) @ mem |> EL ++ YubiPress(pid,tc)\}] \end{alltt}}

\noindent
Finally, the command shown in Figure \ref{fig:yubi-login} represents what happens when the server receives a login request.
This request is accepted if the counter inside the encryption is larger than the last counter stored on the server. 

{\scriptsize\begin{alltt}
:: nil ::
[ \{yubikey -> yubikey ;; 1-1 ;; Server(pid,sid,otc) @ SharedKey(pid,k) @ mem |> EL \},
   -(pid ; nonce ; senc(sid ; tc ; pr, k)), -(otc),  (tc eq (otc + extra)),
  \{yubikey -> yubikey ;; 1-1 ;; Server(pid,sid,tc) @ SharedKey(pid,k) @ mem 
      |> EL ++ Login(pid,sid,tc,senc(sid ; tc ; pr, k)) ++ LoginCounter(pid,otc,tc) \} ]
\end{alltt}}

 \subsection{Formal Specification of YubiHSM in Maude-NPA}\label{sec:APICommandsMNPAYubiHSM}
 \vspace{-1ex}

We consider only the two commands shown in Figures~\ref{fig:yubiHSM-block-encrypt}
and \ref{fig:yubiHSM-aead-generate}.
Each command 
is specified in Maude-NPA as a strand. YubiHSM  makes extensive use of exclusive-or,
denoted by the symbol $*$,
which satisfies the following equations:
\vspace{-1ex}
\begin{align}
x * (y * z) &= (x * y) * z & \mbox{(associativity)}\notag\\
x * y &= y * x & \mbox{(commutativity)} \notag\\
x * null &= x &\mbox{(identity element)}\notag\\
x * x &= null &\mbox{(self-cancellation)} \notag
\end{align}


The YubiHSM command of Figure~\ref{fig:yubiHSM-block-encrypt} is defined as follows. 

{\scriptsize\begin{alltt}
:: nil ::
[ \{YubiHSM -> YubiHSM ;; 1-1 ;; HSM(kh,k) @ mem |> EL \},
   -(kh), -(nonce),
   +(senc(cmode(nonce),k)),
  \{YubiHSM -> YubiHSM ;; 1-1 ;; HSM(kh,k) @ mem |> EL ++ SEnc(kh,nonce) \} ]\end{alltt}}
  
We use two alternative definitions of the YubiHSM command of Figure~\ref{fig:yubiHSM-aead-generate}, one
to represent what happens when the command processes plaintext from the intruder, and another to represent 
what happens when the command processes plaintext from a legitimate principal.  This is possible because,
unlike in the traditional Dolev-Yao model, honest principals communicate with the YubiHSM devices directly,
not through the intruder.  This means that we can represent an honest principal's input data as internal to the
system.   Moreover, in this instance such a representation is necessary, since we are asking whether the intruder
can learn the input data.  We maximize the intruder's advantage, however, by giving it control over the
other input data.

The following strand represents the intruder learning an honest principal's input plaintext data.
We assume the plaintext data is a \sort{Fresh} value.
In this way, we can later ask whether the intruder is able to learn that \sort{Fresh} value.
We use the following macro:
\texttt{aead(n,k,d) = (senc(cmode(n),k) * d) ; mac(d,k)}.

{\scriptsize\begin{alltt}
:: data :: 
[ \{YubiHSM -> YubiHSM ;; 1-1 ;; HSM(kh,k) @ mem |> EL \},
   -(kh), -(nonce),
   +(aead(nonce,k,Fr(data))),
   \{YubiHSM -> YubiHSM ;; 1-1 ;; HSM(kh,k) @ mem
      |> EL ++ GenerateAEAD(Fr(data),aead(nonce,k,Fr(data)))\}]
\end{alltt}}

In the second strand, we replace the \sort{Fresh} value (\texttt{data}) associated to the plaintext data by an input from the intruder.

{\scriptsize\begin{alltt}
:: nil :: 
[ \{YubiHSM -> YubiHSM ;; 1-1 ;; HSM(kh,k) @ mem |> EL \},
   -(data), -(kh), -(nonce),
   +(aead(nonce,k,data)),
   \{YubiHSM -> YubiHSM ;; 1-1 ;; HSM(kh,k) @ mem
      |> EL ++ GenerateAEAD(data,aead(nonce,k,data)) \}]
\end{alltt}}

\vspace {-2em}

\section{Experiments}\label{sec:experiments}
 
We have been able 
to prove secrecy and authentication properties of YubiKey and to find both attacks on YubiHSM:
\begin{enumerate}[(a)]
\item Absence of replay attacks in YubiKey, i.e., there are no two distinct logins that accept the same counter value. 
\item Correspondence between pressing the button on a YubiKey and a successful login. 
In other words, a successful login must have been preceded by a button pressed for the same counter value. 
\item Counter values of YubiKey are different over time, where a successful login invalidates previous \textit{OTP}s. 
\item If the intruder has access to the sever running YubiKey, where the YubiHSM AES keys are generated, then it is able to obtain plaintext in the clear.
\item If the intruder has no access to the server running YubiKey, it can use previous YubiHSM nonces to obtain AES keys.
\end{enumerate}

Table \ref{output-experiment} summarizes the result of the analyses of the YubiKey and YubiHSM APIs specified in Maude-NPA showing the number of generated nodes in each step. 
The notation ``(1)'' represents 
that  the tool  found 1 solution to the question asked by the attack pattern.   
When the number of generated nodes is $0$, the attack pattern is \emph{unreachable}.

\begin{table}
\centering
\begin{tabular}{p{2cm} P{0.7cm} P{0.7cm} P{0.7cm} P{0.7cm} P{0.7cm} P{0.7cm} P{0.7cm}P{0.7cm}P{0.7cm}}
\toprule

\multirow{2}{*}{Attack Pattern} & \multicolumn{7}{c}{Depth} \\ \cmidrule(l){2-10} 
 & 1 & 2 & 3 & 4 & 5 & 6 & 7 & 8 & 9 \\ \midrule
\rowcolor{gray!15}
YubiKey (a) & 4 & 4 & 9 & 21 & 88 & 160 & 0 \\ 
YubiKey (b) & 4 & 7 & 16 & 14 & 2 & 2 & 5 & 0  \\ 
\rowcolor{gray!15}
YubiKey (c) & 4 & 4 & 6 & 18 & 55 & 80 & 0 \\ 
\begin{tabular}[c]{@{}l@{}}YubiKey Login\end{tabular} & 1 & 1 & 2 & 1 & 1 & 1 & 1 & 1 & 1(1)   \\ \midrule
\rowcolor{gray!15}
YubiHSM (d) & 1 & 2 & 3 & 4 & 7 & 13 & 24 & 40 & 76(1) \\ 
YubiHSM (e) & 4 & 6 & 11 & 26(1) &  &  &  \\ \bottomrule
\end{tabular}
\vspace{1mm}
\caption{Output YubiKey and YubiHSM Experiments}
\label{output-experiment} 
\end{table}

Appendix~\ref{app:experiments} provides 
the specific attack patterns. 
All the details on how the attack patterns are specified and which was the output returned by Maude-NPA
are available at 
\url{http://safe-tools.dsic.upv.es/9TCvkSDp}.
The analyses were completely automatic and we obtained finite search graphs for all the attack patterns.
This was achieved thanks to the use of event list expressions within the attack patterns and
the variant-based SMT solving for Lamport clocks. 
Note that Maude-NPA uses a full specification of exclusive-or, an unbounded session model, and
an active Dolev-Yao intruder model.
Moreover, it does not perform any data abstraction such as a bounded number of nonces,
so there are no false positives or negatives.

\section{Related Work}\label{sec:relatedWork}

There is a vast amount of research on the formal analysis of APIs, so in this related work section  we will concentrate on the work that is closest to ours, namely, the formal analysis of the YubiKey and YubiKey-like systems.  
Further related work on APIs and exclusive-or can be found in \cite{ssr2014,ssr2015}.

Besides the work on formalizing and verifying YubiKey that we have already discussed, there has been further work focused on building tools for analyzing policies for YubiKey and YubiKey-like systems.  

In \cite{Yubico01}, Yubico presents some security arguments on their website. An independent analysis was given by blogger Fredrik Bj\"{o}rck in 2009 \cite{Yubico13,Yubico12}, raising issues that Yubico responded to in a subsequent post. Oswald, Richter, et al. \cite{OswaldRP13} analyze the YubiKey, generation 2, for side-channel attacks. They show that non-invasive measurements of the power consumption of the device allow retrieving the AES-key within approximately one hour of access. The authors mentioned  a more recent version of the YubiKey, the YubiKey Neo  which employs a certified smart-card controller that was designed with regard to implementation attacks and is supposed to be more resilient to power consumption analysis.  

K\"unnemann et al. \cite{Kunnemann14} performed a deep analysis of the different properties of  YubiKey, but unlike our analysis using the Maude-NPA tool, they needed to use different lemmas to check some properties that cannot be done automatically by the Tamarin prover, whereas these properties can be checked out in an  automatic way by the Maude-NPA tool. Some properties were not proved due to limited support for exclusive-or.

Mutable global state memory can be used in protocols that provide end-to-end encryption for instant messaging \cite{DBLP:conf/eurosp/Cohn-GordonCDGS17} as well as at the Trusted Platform Module (TPM) \cite{DBLP:conf/csfw/DelauneKRS11} that is a hardware chip designed to enable commodity computers to achieve greater levels of security than is possible in software alone.


 \section{Conclusions}\label{sec:conclusion}
 
 In this paper we have described the analyses of the YubiKey generation 2 and the version 0.9.8 of the YubiHSM
 in Maude-NPA. 
 This allowed us to perform the analysis of these APIs in
a fully-unbounded
 session model making no abstraction  or approximation of fresh values, and
 with no extra assumptions.  
 K{\"u}nnemann and Steel used the Tamarin Prover \cite{Tamarin} 
to prove secrecy and authentication properties of YubiKey and to find the first attack on YubiHSM,
but could not find the second attack on YubiHSM due to the limited use of exclusive-or in the version of Tamarin used in \cite{Kunnemann14}.
We consider our work as complementary to the work of K{\"u}nnemann and Steel.  
 
 
 
 The main challenges involved in modeling and analyzing YubiKey and YubiHSM are: 
(1) handling of Lamport clocks, 
(2) modeling of mutable memory, 
(3) handling of constraints on the ordering of events, and
(4) support for symbolic reasoning modulo exclusive or. 
Very few tools are well equipped to simultaneously handle all of these challenges.

To the best of our knowledge, the most advanced formal modeling and analyses of YubiKey and YubiHSM properties carried out so far is the one in \cite{Kunnemann14}, which uses the Tamarin tool \cite{Tamarin}. Tamarin was well-equipped to handle 
Challenges (1)-(3), but the analysis is incomplete due to the limited support of Tamarin for symbolic reasoning modulo exclusive or in \cite{Kunnemann14}. 
The main goal of this work has been to investigate whether Maude-NPA could complement and extend the formal modeling and analysis results about YubiKey and YubiHSM obtained so far. This is a non-obvious question: on the one hand, Maude-NPA fully supports reasoning modulo exclusive or, so it is well-suited for meeting challenge (4). But on the other hand, previous applications of Maude-NPA have not addressed Challenges (1)-(3). 
The main upshot of the results we present can be summarized as follows: (a) Challenges (2) and (3) can by met by expressing mutable memory and events in terms of \emph{synchronization messages}, a notion used in Maude-NPA to specify protocol compositions \cite{DBLP:journals/corr/SantiagoEMM16}, and (b) Challenge (1) can be met by a
 slight extension of Maude-NPA's current support for equality and disequality constraints \cite{DBLP:conf/birthday/EscobarMMS15}, namely, by adding also support for constraints in
 Presburger Arithmetic. In this way, we show how challenges (1)-(4) 
 can all be met by Maude-NPA, and how these results in formal analyses of YubiKey and YubiHSM that substantially extend previous analyses.    

 What remains to be seen is how generally applicable these tools are to YubiKey and similar APIs.  
 We note that previous work on analyzing API protocols in Maude-NPA did not achieve termination of the search space:
 the IBM CCA API in \cite{ssr2014}
and
the PKCS\#11 in \cite{ssr2015}.
In this work we have been able to achieve termination of many properties thanks to the use of Lamport clocks, mutable memory, 
and event lists.
But more secure API case studies are needed to further test and advance the techniques presented here. 
 
\vspace {-1ex}

\begin{thebibliography}{10}

\bibitem{Maude-NPA}
Maude-{NPA} manual v3.0.
\newblock Available on:
  \url{http://maude.cs.illinois.edu/w/images/d/d5/Maude-NPA_manual_v3.pdf}.

\bibitem{Yubico01}
{Yubico AB. YubiKey Security Evaluation: Discussion of security properties and
  best practices. v2.0.}
\newblock Available on: \url{http://static.
  yubico.com/var/uploads/pdfs/Security_Evaluation_2009- 09-09.pdf}.

\bibitem{Yubico07}
Yubico customer list.
\newblock Available on: \url{http://www.yubico.com/references}.

\bibitem{Yubico105}
{Yubico Inc. YubiHSM 1.0 security advisory 2012-01. 2012.}
\newblock Available on: \url{http://static.yubico.com/var/uploads/pdfs/
  SecurityAdvisory%202012-02-13.pdf}.

\bibitem{FIDO}
{S}pecifications {O}verview, {FIDO} {A}lliance.
\newblock Available on:
  \url{https://fidoalliance.org/specifications/overview/}, December 2015.

\bibitem{Yubico13}
Fredrik Bj{\"o}rck.
\newblock {Security DJ Blog: Increased security for Yubikey}.
\newblock Available on: \url{http://web.archive.org/web/20100203110742/
  http://security.dj/?p=4}, August 2009.

\bibitem{Yubico12}
Fredrik Bj{\"o}rck.
\newblock {Security DJ Blog: Yubikey Security Weaknesses}.
\newblock Available on: \url{ http://web.archive.org/web/20100725005817/
  http://security.dj/?p=154}, February 2009.

\bibitem{DBLP:conf/eurosp/Cohn-GordonCDGS17}
Katriel Cohn{-}Gordon, Cas J.~F. Cremers, Benjamin Dowling, Luke Garratt, and
  Douglas Stebila.
\newblock A formal security analysis of the signal messaging protocol.
\newblock In {\em 2017 {IEEE} European Symposium on Security and Privacy,
  EuroS{\&}P 2017, Paris, France, April 26-28, 2017}, pages 451--466. {IEEE},
  2017.

\bibitem{DBLP:conf/csfw/DelauneKRS11}
St{\'{e}}phanie Delaune, Steve Kremer, Mark~Dermot Ryan, and Graham Steel.
\newblock Formal analysis of protocols based on {TPM} state registers.
\newblock In {\em Proceedings of the 24th {IEEE} Computer Security Foundations
  Symposium, {CSF} 2011, Cernay-la-Ville, France, 27-29 June, 2011}, pages
  66--80. {IEEE} Computer Society, 2011.

\bibitem{DBLP:conf/birthday/EscobarMMS15}
Santiago Escobar, Catherine~A. Meadows, Jos{\'{e}} Meseguer, and Sonia
  Santiago.
\newblock Symbolic protocol analysis with disequality constraints modulo
  equational theories.
\newblock In Chiara Bodei, Gian~Luigi Ferrari, and Corrado Priami, editors,
  {\em Programming Languages with Applications to Biology and Security - Essays
  Dedicated to Pierpaolo Degano on the Occasion of His 65th Birthday}, volume
  9465 of {\em Lecture Notes in Computer Science}, pages 238--261. Springer,
  2015.

\bibitem{strands}
F.~J.~T. Fabrega, J.~C. Herzog, and J.~D. Guttman.
\newblock Strand spaces: why is a security protocol correct?
\newblock In {\em Proceedings. 1998 IEEE Symposium on Security and Privacy
  (Cat. No.98CB36186)}, pages 160--171, May 1998.

\bibitem{ssr2014}
Antonio Gonz{\'{a}}lez{-}Burgue{\~{n}}o, Sonia Santiago, Santiago Escobar,
  Catherine~A. Meadows, and Jos{\'{e}} Meseguer.
\newblock {Analysis of the {IBM} {CCA} Security {API} Protocols in Maude-NPA}.
\newblock In Liqun Chen and Chris~J. Mitchell, editors, {\em {Security
  Standardisation Research - First International Conference, {SSR} 2014,
  London, UK, December 16-17, 2014. Proceedings}}, volume 8893 of {\em Lecture
  Notes in Computer Science}, pages 111--130. Springer, 2014.

\bibitem{ssr2015}
Antonio Gonz{\'{a}}lez{-}Burgue{\~{n}}o, Sonia Santiago, Santiago Escobar,
  Catherine~A. Meadows, and Jos{\'{e}} Meseguer.
\newblock {Analysis of the PKCS{\#}11 {API} Using the Maude-NPA Tool}.
\newblock In Liqun Chen and Shin'ichiro Matsuo, editors, {\em Security
  Standardisation Research - Second International Conference, {SSR} 2015,
  Tokyo, Japan, December 15-16, 2015, Proceedings}, volume 9497 of {\em Lecture
  Notes in Computer Science}, pages 86--106. Springer, 2015.

\bibitem{Kunnemann14}
Robert K{\"u}nnemann.
\newblock {\em {Foundations for analyzing security APIs in the symbolic and
  computational model. Available on:
  \url{https://tel.archives-ouvertes.fr/tel-00942459/file/Kunnemann2014.pdf}}}.
\newblock Theses, {{\'E}cole normale sup{\'e}rieure de Cachan - ENS Cachan},
  January 2014.

\bibitem{Kunnemann12}
Robert K{\"u}nnemann and Graham Steel.
\newblock {Y}ubi{S}ecure? formal security analysis results for the {Y}ubikey
  and {Y}ubi{HSM}.
\newblock In Audun J{\o}sang, Pierangela Samarati, and Marinella Petrocchi,
  editors, {\em {R}evised {S}elected {P}apers of the 8th {W}orkshop on
  {S}ecurity and {T}rust {M}anagement ({STM}'12)}, volume 7783 of {\em Lecture
  Notes in Computer Science}, pages 257--272, Pisa, Italy, September 2012.
  Springer.

\bibitem{Lamport:1978:TCO:359545.359563}
Leslie Lamport.
\newblock Time, clocks, and the ordering of events in a distributed system.
\newblock {\em Commun. ACM}, 21(7):558--565, July 1978.

\bibitem{Tamarin}
Simon Meier, Benedikt Schmidt, Cas Cremers, and David~A. Basin.
\newblock The {TAMARIN} prover for the symbolic analysis of security protocols.
\newblock In Natasha Sharygina and Helmut Veith, editors, {\em Computer Aided
  Verification - 25th International Conference, {CAV} 2013, Saint Petersburg,
  Russia, July 13-19, 2013. Proceedings}, volume 8044 of {\em Lecture Notes in
  Computer Science}, pages 696--701. Springer, 2013.

\bibitem{Merkel09}
Dirk Merkel.
\newblock {Linux Journal: Yubikey One-Time Password Authentication}.
\newblock Available on:
  \url{http://dl.acm.org/citation.cfm?id=1502508.1502509}, January 2009.

\bibitem{MeseguerSCP18}
Jos{\'{e}} Meseguer.
\newblock Variant-based satisfiability in initial algebras.
\newblock {\em Sci. Comput. Program.}, 154:3--41, 2018.

\bibitem{Nieuwenhuis:2006:SSS:1217856.1217859}
Robert Nieuwenhuis, Albert Oliveras, and Cesare Tinelli.
\newblock Solving SAT and SAT modulo theories: From an abstract
  Davis--Putnam--Logemann--Loveland Procedure to DPLL(t).
\newblock {\em J. ACM}, 53(6):937--977, November 2006.

\bibitem{OswaldRP13}
David Oswald, Bastian Richter, and Christof Paar.
\newblock Side-channel attacks on the Yubikey 2 one-time password generator.
\newblock In Salvatore~J. Stolfo, Angelos Stavrou, and Charles~V. Wright,
  editors, {\em Research in Attacks, Intrusions, and Defenses - 16th
  International Symposium, {RAID} 2013, Rodney Bay, St. Lucia, October 23-25,
  2013. Proceedings}, volume 8145 of {\em Lecture Notes in Computer Science},
  pages 204--222. Springer, 2013.

\bibitem{DBLP:journals/corr/SantiagoEMM16}
Sonia Santiago, Santiago Escobar, Catherine~A. Meadows, and Jos{\'{e}}
  Meseguer.
\newblock Effective sequential protocol composition in Maude-NPA.
\newblock {\em CoRR}, abs/1603.00087, 2016.

\bibitem{Vamanu12}
L.~Vamanu.
\newblock {Formal analysis of Yubikey. Master's thesis, \'{E}cole normale
  sup{\'{e}}rieure de Cachan (August 2011)}.
\newblock Available on:
  \url{http://n.ethz.ch/~lvamanu/download/YubiKeyAnalysis.pdf}.

\bibitem{Yubico-manual}
Yubico.
\newblock {The YubiKey Manual. Usage, configuration and introduction of basic
  concepts. Version: 3.4}.
\newblock Available on: \url{https://goo.gl/zk5fbK}.

\bibitem{YubiHMS-manual}
Yubico.
\newblock {YubiHSM Manual v1.5}.
\newblock Available on:
  \url{https://www.yubico.com/wp-content/uploads/2015/04/YubiHSM-Manual_1_5_0.pdf}.

\end{thebibliography}

\vspace {-1ex}

\appendix

\vspace {-0.5ex}
\section{Experiments}\label{app:experiments} 
\vspace {-0.5ex}

\vspace {-0.5ex}
 \subsection{Absence of replay attacks in YubiKey}
\vspace {-0.5ex}
 
 The first property to prove is  the absence of replay attacks, where there are no two distinct logins that accept the same counter value. 
 This property is specified in Maude-NPA as follows:
 
\vspace {-1ex}
 {\scriptsize\begin{alltt}
eq ATTACK-STATE(1)
= :: nil ::
  [ nil, \{yubikey -> yubikey ;; 1-1 ;; mem |> 
         EL1 ++ Login(pid,sid,counter2,X) ++ EL2 ++ Login(pid,sid,counter2,Y) ++ EL3 \} | nil ]
  || empty || nil || nil || nil [nonexec] .
 	\end{alltt}}
\vspace {-1ex} 	

\noindent 
The output of this attack shows that an initial state cannot be reached  and,  since there is a finite search space, this property  is secure under these circumstances. The number of generated states  of this attack pattern  is shown at the ``YubiKey (a)'' row of Table \ref{output-experiment}. 

\subsection{Correspondence in YubiKey}
\vspace {-1ex}

The second property represents a 
correspondence between pressing the button on a YubiKey and a successful login. A successful login must have been preceded by a button pressed for the same counter value. 
That is, we are searching for 
 \texttt{EL1++ Login(pid,sid,1 + c4,Y) ++ EL2}
with the condition that
\texttt{EL1}  is not of the form  \texttt{EL3 ++ Yubipress(pid,1 + c4) ++ EL4},
which is defined by the never pattern below.
This property is specified in Maude-NPA as follows:

\vspace {-1ex}
{\scriptsize\begin{alltt}
eq ATTACK-STATE(2)
 = :: nil ::
[ nil, \{yubikey -> yubikey ;; 1-1 ;; mem |> EL1 ++ Login(pid,sid,1 + c4,Y) ++ EL2 \} | nil ]
   || empty || nil || nil
   || never (
   :: nil ::
   [ nil | \{yubikey -> yubikey ;; 1-1 ;; mem |> 
          EL1a ++ YubiPress(pid,1 + c4) ++ EL1b ++ Login(pid,sid,1 + c4,Y) ++ EL2\}, nil ] 
     & S:StrandSet || K:IntruderKnowledge) [nonexec] .
\end{alltt}} 
\vspace {-1ex}

\noindent
 The output of this attack shows that an initial state cannot be reached  and,  since there is a finite search space, this property is secure under these circumstances. The number of generated states  of this attack pattern  is shown at the ``YubiKey (b)'' row of Table \ref{output-experiment}.

%

\subsection{Counter value control in YubiKey} 	 
\vspace {-1ex}
 	 
 The last property of YubiKey represents the  control of the counter values, that 
 requires that if two logins occur with the same pid, the counter of the first one 
 should be less than the counter of the second one.

	 Note that the verification of this property in Tamarin \cite{Kunnemann14} requires additional user-defined lemmas.
This property is specified in Maude-NPA as follows:
 
\vspace {-1ex}
 {\scriptsize\begin{alltt}
eq ATTACK-STATE(3)
= :: nil ::
     [ nil, \{yubikey -> yubikey ;; 1-1 ;; mem |>
        EL1 ++ LoginCounter(pid,c1,c2 + c4) ++ EL2 ++ LoginCounter(pid,c3,c4) ++ EL3 \} | nil ]
  || empty || nil || nil || nil [nonexec] .
 	\end{alltt}}
\vspace {-1ex}

\noindent 	
The output of this attack shows that an initial state cannot be reached  and,  since there is a finite search space, it is secure under these circumstances. The number of generated states  of this attack pattern  is shown at the ``YubiKey (c)'' row of Table \ref{output-experiment}.

%
\subsection{A Regular Execution in YubiKey}
\vspace {-1ex}

 	 Finally, we specified an attack pattern that represents a regular login sequence of the YubiKey API. The specification is as follows:
 	 
\vspace {-1ex}
{\scriptsize\begin{alltt}
eq ATTACK-STATE(0) =
:: nil :: 
[nil, \{yubikey -> yubikey ;; 1-1 ;; mem |> 
       Init(pid,k) ++ ExtendedInit(pid,sid,k) ++ YubiPress(rpid,1) ++ 
       Login(pid, sid, (1 + 1),senc(sid ; (1 + 1) ; npr, k)) ++ LoginCounter(pid,1,1 + 1) \} 
       | nil ] || empty || nil || nil || nil [nonexec] .
\end{alltt}} 	
\vspace {-1ex}

\noindent
The output of the execution of this attack pattern in Maude-NPA  shows  that Maude-NPA finds an initial state, 
showing that it is possible to perform a regular login. The number of generated states  of this attack pattern  is shown at the ``YubiKey Regular Execution'' row of Table \ref{output-experiment}.

%
\subsection{First Attack on YubiHSM} 	  
\vspace {-1ex}

In this attack, the intruder has access to the server where YubiKey AES keys are generated.
The intruder can extract the plaintext sent within an AEAD if it was able to listen to a previous 
call to the command of 
Figure~\ref{fig:yubiHSM-block-encrypt}.
The Maude-NPA specification is as follows:

\vspace {-1ex}
{\scriptsize\begin{alltt}
eq ATTACK-STATE(1) =
:: kh, k ::
[nil, +(Fr(kh:Fresh)),
      \{YubiHSM -> YubiHSM ;; 1-1 ;; HSM(Fr(kh),Fr(k)) @ YSM-AEAD(Fr(kh)) 
      |>  MasterKey(Fr(k))\} | nil ]
 &
::: data :: 
[nil, \{YubiHSM -> YubiHSM ;; 1-1 ;; HSM(Fr(kh),Fr(k)) @ YSM-AEAD(Fr(kh)) 
      |> MasterKey(Fr(k)) ++ SEnc(Fr(kh),cmode(nonce))\},
       -(Fr(kh)), -(Fr(nonce)),
       +(aead(cmode(nonce),Fr(k),Fr(data))) | nil  ]
|| Fr(data) inI || nil || nil || nil [nonexec] .

\end{alltt}} 	  
\vspace {-2ex}

\noindent
The output of the execution of this attack in Maude-NPA  shows that Maude-NPA finds an initial state, proving the existence of this attack. The number of generated states  of this attack pattern  is shown at the ``YubiHSM (d)'' row of Table \ref{output-experiment}.
%
%
The initial state found by the tool shows the attack

\vspace {-1ex}
{\scriptsize\begin{alltt}
:: x1,x3 ::
[ nil | 
   +(Fr(x1)), 
   \{YubiHSM -> YubiHSM ;; 1-1 ;; YSM-AEAD(Fr(x1)) @ HSM(Fr(x1), Fr(x3)) 
   |> MasterKey(Fr(x3))\}, nil] ) &
:: nil ::
[ nil | \{YubiHSM -> YubiHSM ;; 1-1 ;; YSM-AEAD(Fr(x1)) @ HSM(Fr(x1), Fr(x3)) 
|> MasterKey(Fr(x3))\}, 
   -(Fr(x1)), -(x2), 
   +(senc(cmode(x2), Fr(x3))), 
   \{YubiHSM -> YubiHSM ;; 1-1 ;; YSM-AEAD(Fr(x1)) @ HSM(Fr(x1), Fr(x3))  
   |> MasterKey(Fr(x3)) ++ SEnc(Fr(x1), x2)\}, nil] &
:: x0 ::
[ nil | \{YubiHSM -> YubiHSM ;; 1-1 ;; YSM-AEAD(Fr(x1)) @ HSM(Fr(x1), Fr(x3)) 
|> MasterKey(Fr(x3)) ++ SEnc(Fr(x1), x2)\}, 
   -(Fr(x1)), -(x2), 
   +((Fr(x0) * senc(cmode(x2), Fr(x3))) ; mac(Fr(x0), Fr(x3))), nil]  & 
:: nil ::
[ nil |
   -((Fr(x0) * senc(cmode(x2), Fr(x3))) ; mac(Fr(x0), Fr(x3))), 
   +(Fr(x0) * senc(cmode(x2), Fr(x3))), nil]  & 
:: nil ::
[ nil | 
   -(senc(cmode(x2), Fr(x3))), -(Fr(x0) * senc(x2, Fr(x3))), 
   +(Fr(x0)), nil]  
\end{alltt}}
\vspace {-1ex}

\vspace {-2ex}
\subsection{Second Attack on YubiHSM} 	  

In this attack \cite{Yubico105}, the YubiHSM command
of Figure~\ref{fig:yubiHSM-aead-generate}
can be used to  decrypt a previously generated AEAD. 
If an intruder:
 \begin{enumerate}
  \item can acquire a previously generated AEAD, together with the nonce value used when
generating the AEAD, and
\item can use a YubiHSM with the same key handle that generated the first AEAD to generate a
second AEAD with arbitrary nonce and plaintext,
\end{enumerate}

\noindent then the intruder will be able to decrypt the plaintext from the first AEAD using the YubiHSM. Furthermore, the intruder can extract any symmetric encryption  by calling the AEAD generate API call with data = 0. The Maude-NPA specification is as follows:

{\scriptsize\begin{alltt}
eq ATTACK-STATE(2) =
:: kh, k ::
[nil, +(Fr(kh)),
      \{YubiHSM -> YubiHSM ;; 1-1 ;; HSM(Fr(kh),Fr(k)) @ YSM-AEAD(Fr(kh)) |> MasterKey(Fr(k)) \} 
      | nil ]
 &
:: nil :: 
[nil, \{YubiHSM -> YubiHSM ;; 1-1 ;; HSM(Fr(kh),Fr(k)) @ YSM-AEAD(Fr(kh)) |> MasterKey(Fr(k)) \},
       -(null), -(Fr(kh)), -(nonce),
       +(aead(nonce,Fr(k),null)) | nil ]
|| senc(nonce,Fr(k))  inI || nil || nil || nil [nonexec] .
\end{alltt}} 
\vspace{-1ex}
 
\noindent 
The output of the execution of this attack in Maude-NPA  shows that Maude-NPA finds an initial state, proving the existence of this attack on the YubiHSM API. The number of generated states  of this attack pattern  is shown at the ``YubiHSM (e)'' row of Table \ref{output-experiment}.  
%
%
The initial state found by the tool shows the attack

\vspace{-1ex}
{\scriptsize\begin{alltt}
:: x2,x0 :: [ nil | 
   +(Fr(x0)), 
   \{YubiHSM -> YubiHSM ;; 1-1 ;; YSM-AEAD(Fr(x0)) @ HSM(Fr(x0), Fr(x2)) |> MasterKey(Fr(x2)) \}, 
   nil]  &
:: nil :: [ nil | 
   \{YubiHSM -> YubiHSM ;; 1-1 ;; YSM-AEAD(Fr(x0)) @ HSM(Fr(x0), Fr(x2)) |> MasterKey(Fr(x2)) \}, 
   -(null), -(Fr(x0)), -(x1), 
   +(senc(cmode(x1), Fr(x2)) ; mac(null, Fr(x2))), nil] &
:: nil :: [ nil | 
   -(senc(cmode(x1), Fr(x2)) ; mac(null, Fr(x2))), 
   +(senc(cmode(x1), Fr(x2))), nil]  
\end{alltt}}
\vspace{-1ex}

\noindent
According to the experiments reported in \cite{Kunnemann12},
it appears that Tamarin cannot find the second attack due to the limited use of exclusive-or.

\end{document}